\documentclass[aps,prl, reprint,superscriptaddress, preprintnumbers]{revtex4-1}
\usepackage[english]{babel}
\usepackage{amsmath,amscd}
\usepackage{wasysym}
\usepackage{graphicx}
\usepackage{caption}
\usepackage{tikz}
\usepackage{amssymb}
\pdfoutput=1

\def\be{\begin{equation}}
\def\ee{\end{equation}}
\def\bea{\begin{eqnarray}}
\def\eea{\end{eqnarray}}
\def\pd{\partial}
\def\a{\alpha}
\def\b{\beta}

\def\m{\mu}
\def\n{\nu}

\def\l{\lambda}

\def\r{\rho}

\def\s{\sigma}

\def\bma{\begin{pmatrix}}
\def\ema{\end{pmatrix}}

\def\bi{\begin{itemize}}
\def\ei{\end{itemize}}

\begin{document} 
	\preprint{IFT-UAM/CSIC-16-038}
	\preprint{FTUAM-16-14}
	\preprint{FTI/UCM-16-27}
	
\title{\boldmath  A note on the Gauge Symmetries of Unimodular Gravity.}
\author{Enrique \'Alvarez}
\email{enrique.alvarez@uam.es}
\affiliation{ Instituto de F\'{\i}sica Te\'orica, IFT-UAM/CSIC, Universidad Aut\'onoma, 28049 Madrid, Spain}
\affiliation{Departamento de F\'{\i}sica Te\'orica, Universidad Aut\'onoma de Madrid, 28049 Madrid, Spain}
\author{Sergio Gonz\'alez-Mart\'{\i}n}
\email{sergio.gonzalez.martin@csic.es}
\affiliation{ Instituto de F\'{\i}sica Te\'orica, IFT-UAM/CSIC, Universidad Aut\'onoma, 28049 Madrid, Spain}
\affiliation{Departamento de F\'{\i}sica Te\'orica, Universidad Aut\'onoma de Madrid, 28049 Madrid, Spain}
\author{Carmelo P. Mart\'{\i}n}
\email{carmelop@fis.ucm.es}
\affiliation{Universidad Complutense de Madrid (UCM), Departamento de F\'isica Te\'orica I,
Facultad de Ciencias F\'{\i}sicas, Av. Complutense S/N (Ciudad Univ.),
28040 Madrid, Spain}

\begin{abstract}
\centering The symmetries of Unimodular Gravity are clarified somewhat.
\end{abstract}

\maketitle

\section{Introduction}
Unimodular gravity (UG) is a truncation of General Relativity (GR) in the sense that only unimodular metrics (i.e. those with unit determinant) are considered. A recent review is \cite{Alvarez} where some early references can be found. It is remarkable that Einstein himself proposed a closely related theory in 1919 \cite{Einstein}.
\par
The theory can be (and it is technically convenient) formulated in such a way that it has an added Weyl invariance by writing
\be\label{Ur}
\hat{g}_{\m\n}\equiv \left(T_U g\right)_{\m\n}\equiv  \left|g\right|^{-{1\over n}}~g_{\m\n}
\ee
(where $g\equiv \text{det}~g_{\a\b}$).
The reason is that then the variations $\delta g_{\a\b}$ are unconstrained, whereas the variations of the unimodular metric have got to be traceless
\be
\hat{g}^{\a\b}\delta \hat{g}_{\a\b}=0
\ee
We shall denote the mapping
\be
\text{UR}:~g_{\m\n}\rightarrow \hat{g}_{\a\b}
\ee
as {\em unimodular reduction}. It is not invertible, since there is no way to reconstruct $g_{\a\b}$ from its unimodular reduction $\hat{g}_{\a\b}$.
On the other hand, once we restrict the theory to unimodular metrics, the ensuing theory (UG) is not invariant under the full diffeomorphism group of the manifold. $\text{Diff}(M)$, but only under the subgroup that preserves the unimodularity condition, which we have dubbed $\text{TDiff}(M)$. This is essentially what mathematicians call  the volume preserving subgroup \cite{Dusa}. It has been pointed out that this symmetry is enough to kill the three unwanted polarizations when defining the massless theory from a massive theory in flat space \cite{vanderBij}.
At any rate, under unimodular reduction Einstein-Hilbert action gets transformed into

\bea
\text{UR}:~ &&S_{GR}\equiv  - M_P^{n-2}\int d^n x ~\sqrt{|g|}~R\left[g_{\a\b}\right] \longrightarrow\nonumber \\ \longrightarrow \quad &&S_{UG}\equiv - M_P^{n-2}\int d^n x ~R\left[\hat{g}_{\a\b}\right]
\eea
and the unimodular action in terms of unconstrained variables reads
\be
S_{UG}=-M_P^{n-2}\int d^n x ~ |g|^{1\over n}~\left(R+{(n-1)(n-2)\over 4 n^2}{\nabla_\m g\nabla^\m g\over g^2}\right)
\ee
Once here one can never go back to the Einstein frame as this action is Weyl invariant.

In terms of this unconstrained metric, the equations of motion (EM) are given  by the manifestly traceless expression \cite{AlvarezGMHVM}\bea
&&R_{\m\n}-{1\over n}~g_{\m\n}=\Theta_{\m\n}\nonumber\\
&&\Theta_{\m\n}\equiv {(n-2)(2n-1)\over 4 n^2}\left({\nabla_\m g \nabla_\n g\over g^2}-{1\over n}{(\nabla g)^2\over g^2}~g_{\m\n}\right)-\nonumber\\
&&-{n-2\over 2n}\left({\nabla_\m \nabla_\n g \over g}-{1\over n}{\nabla^2 g\over g} g_{\m\n}\right)\label{UGEM}
\eea

 The explicit presence of the determinant of the metric, $g$ clearly indicates the EM are not $\text{Diff}$ invariant. The covariant derivative acting on $g(x)$ is defined as
 \bea
 &&\nabla_\m g(x)\equiv \pd_\m g(x)\nonumber\\
 &&\nabla_\s\nabla_\m g(x)\equiv \pd_\s\pd_\m g(x)-\Gamma^\l_{\s\m}\pd_\l g(x)
 \eea
 and has bizarre transformation properties.

Now given the fact that the EM are Weyl invariant, we can always transform to 
\be
\hat{g}=1
\ee
where the EM simply read
\be
\hat{R}_{\m\n}=\frac{1}{n}\hat{R} \hat{g}_{\m\n}
\ee
The solution of these equations are by definition \textit{Einstein spaces} \cite{Besse}. The Bianchi identities in the absence of torsion do imply then $\nabla_\m R=0$.
\par
Given an unimodular Einstein space, $\hat{g}_{\m\n}$, all its Weyl rescalings
\be
g_{\m\n}\equiv \Omega^2(x)~\hat{g}_{\m\n}
\ee
are also solutions of the equations [\ref{UGEM}]. They span a {\em Weyl orbit} of solutions. In four dimensions it is well known that the necessary and sufficient condition \cite{Kozameh} for a space to be conformally Einstein is for it to be Bach-flat
\be
B_{\m\n}\equiv \nabla^\a \nabla^\b W_{\a\m\n\b}-{1\over 2} R^{\a\b}~W_{\a\m\n\b}=0
\ee
where $W_{\m\n\r\s}$ is the Weyl tensor. We are not aware of a similar statement in arbitrary dimension.

\par
The full symmetry group of this action is quite large though, incorporating Weyl transformations of the metric. This means that in the process of unimodular reduction of Einstein-Hilbert the symmetry group changes, namely 
\be
\text{UR}:\quad\text{Diff}(M)\longrightarrow\text{TDiff}(M)\ltimes \text{Weyl}(M)
\ee
Let us examine this process of symmetry reduction  in more detail. We shall be cavalier about domains of definition of the transformations, and all of our reasoning will be purely local.

\section{Tdiff invariance of the unimodular action}
It is not immediately obvious in which reference systems are the EM [\ref{UGEM}] valid. 
\par

Let us first start with the analysis of the already mentioned change of the symmetry group in the process of unimodular reduction.\par
We can represent a linearized element of $\text{Diff}_0(M)$ (the subgroup of  $\text{Diff}(M)$ connected with the identity)  as
\be
x\rightarrow x^\prime\equiv x+\xi
\ee
The corresponding jacobian matrix is
\be
J^\a_{\b^\prime}(x)\equiv~{\pd x^\a\over \pd x^{\b^\prime}}
\ee
and its determinant will be denoted by the letter $J$.

\par
The determinant of the metric then transforms as
\be
g(x)\rightarrow g^\xi\left(x+\xi\right)=J^2(x)~ g(x)
\ee

And for the case of a volume preserving diffeomorphism, it is transverse in the sense that
\be
\pd_\l \xi_T^\l=0
\ee
and the jacobian matrix is itself unimodular
\be
J_T=1
\ee
Let us examine what happens with the action of $\text{TDiff}_0\ltimes \text{Weyl}(M)$. Clearly
\be
g_{\m\n}^\xi(x)\equiv J_\m^\a J_\n^\b g_{\a\b}(x-\xi)
\ee
and consequently
\be
g_{\m\n}^{\xi \Omega}\equiv \Omega^2(x)~ J_\m^\a J_\n^\b g_{\a\b}(x-\xi)
\ee
On the other hand, the other way around
\be
g_{\m\n}^{\Omega\xi}(x)=\Omega^2(x-\xi)~ J_\m^\a J_\n^\b g_{\a\b}(x-\xi)
\ee
This corresponds to the non-commutativity of the diagram
\be
\begin{CD}
g_{\m\n} @>TDiff>> g^\xi_{\m\n}\\
@VVWeylV @VVWeylV\\
g_{\m\n }^\Omega @>TDiff>> g_{\m\n}^{\xi\Omega}\neq g_{\m\n}^{\Omega\xi}
\end{CD}
\ee
 thus being the reason why the symmetry group is a semidirect product.
 \newline
 
 We can move now to answer the question of the validity of the EM of UG.\par The two possible paths when going from GR to UG are shown in the following diagram.
 
 \be
\begin{CD}
GR @>Diff>> GR\\
@VVURV @VVURV\\
UG @>Diff>> UG
\end{CD}
\ee
 The rightmost path correspond to, first perform a \text{Diff}

\be
g^\xi_{\m\n}(x)\equiv \left(T_\xi g\right)_{\m\n}(x)\equiv J_\m^\a(x-\xi) J_\n^\n(x-\xi) g_{\a\b}(x-\xi)
\ee 
and unimodularly reduce afterwards. The corresponding unimodular metric is then
\be
\left(T_U T_\xi g\right)_{\m\n}(x)=J^{-{2\over n}}(x)~g^{-{1\over n}}(x)~\left(T_\xi g\right)_{\m\n}
\ee

Let us now perform an arbitrary diffeomorphism after unimodular reduction (corresponding to the left path in the diagram). The result is
\be
\left(T_\xi T_U g\right)_{\a\b}(x)= J^{-{2\over n}}(x-\xi) g^{-{1\over n}}(x-\xi)~\left(T_\xi g\right)_{\a\b}
\ee
This means again that the diagram above is not commutative.

Indeed, we find particularly clarifying to examine what happens in this latter case \par If we perform a \text{Diff} in \eqref{UGEM} the determinant $g(x)$ transforms as

\bea
\nabla_{\l^\prime} g^\xi(x^\prime)&&=J^\a_{\l^\prime} \nabla_\a \left(J^2(x)~g(x)\right)=\nonumber\\
&&=~J^\a_{\l^\prime}\left(J^2~\nabla_\a g(x)+2 g(x)~J(x)\nabla_\a J\right)
\eea 
this conveys the fact that the first monomial in the EM transforms as
\begin{widetext}
\bea
{\nabla_{\m^\prime} g^\xi(x^\prime)\nabla_{\n^\prime} g^\xi(x^\prime)\over g^\prime(x^\prime)^2}&&={J_{\m^\prime}^\a J_{\n^\prime}^\b\over J^4 g^2}\left(\nabla_\a g J^2+ 2 g J \nabla_\a J\right)\left( J^2 \nabla_\b g+ 2 g J \nabla_\b J\right)=\nonumber\\
&&=J_{\m^\prime}^\a J_{\n^\prime}^\b \bigg\{{\nabla_\a g\nabla_\b g\over g^2}+2  {\nabla_\a J \nabla_\b g+\nabla_\a g\nabla_\b J\over J g}+4 {\nabla_\a J\nabla_\b J\over J^2}\bigg\}
\eea 
and its trace, which is the one subtracted from it in \eqref{UGEM}, is just
\be
\left({\nabla_\a g \over g}+ 2  {\nabla_\a J\over J}\right)^2
\ee
The second monomial transforms in turn as

\bea 
{\nabla_{\m^\prime}\nabla_{\n^\prime} g^\xi\left(x+\xi\right)\over g^\xi(x+\xi)}&&={J_{\m^{\prime}}^\r~J^\a_{\n^\prime}\nabla_\r\over J^2 g}\left(J^2~\nabla_\a g(x)+2 g(x)~J(x)\nabla_\a J\right)=\nonumber\\
&&=J_{\m^{\prime}}^\r~J^\a_{\n^\prime}\bigg\{2{\nabla_\r J\over J}{\nabla_\a g\over g}+{\nabla_\r\nabla_\a g\over g}+ 2 {\nabla_\r g\over g}{\nabla_\a J\over J}+2{\nabla_\a J\over J}{\nabla_\r J\over J}+2{\nabla_\r\nabla_\a J\over J} \bigg\}
\eea 
being its trace now
\be
4{\nabla_\a J\nabla^\a g\over g J}+ {\nabla^2 g\over  g}+2 {\nabla_\a J\nabla^\a J\over J^2}+2~{\nabla^2 J\over J}
\ee
\end{widetext}
\section{Conclusions}
When performing a general $\text{Diff}_0(M)$ transformation in the unimodular EM the extra terms generated are
\bea
&& EM\left[g_{\m\n}^\xi\right]_{\a^\prime\b^\prime}=J_{\a'}^{\a} J_{\b'}^\b\bigg\{ EM\left[g_{\m\n}\right]_{\a\b}+\nonumber\\
&&+{n-2\over 2n}\left(  {1\over n}{\nabla_\a J \nabla_\b g+\nabla_\a g\nabla_\b J\over J g}+2{1-n\over n} {\nabla_\a J\nabla_\b J\over J^2}+\right.\nonumber\\
&&\left.
+2{\nabla_\a\nabla_\b J\over J}\right)-\\
&&-{n-2\over n^2}
\left( {1\over n}{\nabla_\m J\nabla^\m g\over g J}+{1-n\over n} {\nabla_\m J\nabla^\m J\over J^2}+{\nabla^2 J\over J}
\right)~g_{\a\b}\bigg\}\nonumber\\
\eea

To be specific: the fact that a given metric $g_{\m\n}$ is a solution of the unimodular equations of motion does not imply that it remains a solution after an arbitrary diffeomorphism $\xi\in \text{Diff}_0(M)$ unless of course this happens to be transverse, $\xi\in\text{TDiff}_0(M)$. Certainly there is no problem with performing the \text{Diff} before the unimodular reduction, since GR is invariant as  shown in the last diagram.
\par

In other words, the assertion that a given metric is a solution of the UG equations of
motion is not Diff(M) invariant, but only TDiff(M) $\ltimes $Weyl(M) invariant.
Is there a coordinate system which is not attainable through a symmetry transformation?
It could be thought that there is none, by the following argument. An arbitrary diffeomorphism acts as

\be
g_{\m\n}^\xi(x)\equiv J_\m^\a J_\n^\b g_{\a\b}(x-\xi)
\ee which has the same number of parameters as  the action of a volume preserving diffeomorphism composed with a Weyl transformation. However, there is a subtlety here, since one should have solutions to the equation

\be 
\dfrac{J^\a_\m}{J}=\dfrac{\partial y^\a}{\partial x^\m}
\ee and this is possible only when

\be 
\partial_\n\left(\dfrac{J^\a_\m}{J}\right)=\partial_\m\left(\dfrac{J^\a_\n}{J}\right) \label{integration}
\ee which will not be, in general, true.

In the appendix we work out a simple example to illustrate this fact.

\section{Acknowledgments}
 The origin of this work lies in a question by Gerardus 't Hooft at the \textit{Shapes of Gravity} meeting held in Nijmegen, 2016. It has been partially supported by the European Union FP7 ITN INVISIBLES (Marie Curie Actions, PITN- GA-2011- 289442)and (HPRN-CT-200-00148); COST action MP1405 (Quantum Structure of Spacetime), COST action MP1210 (The String Theory Universe) as well as by FPA2012-31880 (MICINN, Spain)), FPA2014-54154-P (MICINN, Spain), and S2009ESP-1473 (CA Madrid). This project has received funding from the European Union' s Horizon 2020 research and innovation programme under the Marie Sklodowska-Curie grant agreement No 690575. This project has also received funding from the European Union' s Horizon 2020 research and innovation programme under the Marie Sklodowska-Curie grant agreement No 674896.
The authors acknowledge the support of the Spanish MINECO {\em Centro de Excelencia Severo Ochoa} Programme under grant SEV-2012-0249.

\appendix
\section{Any metric can be made unimodular by a diffeomorphism}
All we have to do is to find a solution of the equation
\be
J(x)={1\over g(x)^2}
\ee
At the linear level (algebra)
\be
{\pd_\m\xi^\m}={1\over g(x)^2}-1
\ee
and this is trivially solved in a formal way by
\be
\xi^\m(x)=\pd^\m~\Box^{-1}~\left({1\over g(x)^2}-1\right)
\ee
whose solution is unique under essentially the same conditions as the corresponding solution of the wave equation \cite{Friedlander}.
\section{Flat space in cylindrical coordinates is not a solution of UG}
Let us consider to be specific, the ordinary three-dimensional euclidean space $\mathbb{R}^3$ in cylindrical coordinates:
\be
ds^2=dr^2+r^2 d\theta^2+ dz^2
\ee
A simple calculation, taking into account that
\bea
&&\Gamma^r_{\phi\phi}=-r\nonumber\\
&&\Gamma^\phi_{\phi r}={1\over r}
\eea
yields
\be
\Theta_{\m\n}={1\over 27 r^2}\begin{pmatrix}-7&0 &0\\0&8 r^2&0\\0&0&-1\end{pmatrix}\neq 0
\ee

That is, flat space in cylindrical coordinates is not a solution of the unimodular equations of motion. This happens of course because the transformation from cartesian to cylindrical coordinates does not belong to TDiff($\mathbb{R}^3$).\newline
In terms of the integrability conditions \eqref{integration}, it is clear that there is no integrating factor  because the jacobian from cartesian coordinates read
\be
J^\a_\m\equiv\begin{pmatrix}\cos~\theta&\sin~\theta&0\\-r\sin~\theta&r\cos~\theta&0\\0&0&1\end{pmatrix}
\ee
whose determinant is
\be
J=r
\ee
and the integrability conditions fail here because, for example
\be
{\pd\over \pd \theta}\left({J^x_r\over J}\right)=-{\cos~\theta\over r^2}\neq {\pd\over \pd r}\left({J^x_\theta\over J}\right)=0
\ee

\end{document}